\documentstyle[preprint,aps]{revtex}
\begin{document}
\draft
\title{Hysteresis in the Mott Transition between Plasma and Insulating Gas}

\author{D.W. Snoke and J.D. Crawford}

\address{Department of Physics and Astrophysics,
University of Pittsburgh\\
3941 O'Hara St., Pittsburgh, PA 15260}

\maketitle

\begin{abstract}
We show that hysteresis can occur in the transition between a neutral plasma
and
the insulating gas consisting of neutral pairs bound by Coulomb attraction.
Since the transition depends sensitively on the screening length in the plasma,
regions of bistability occur in density--temperature phase space.  We present
numerical results which indicate where these regions occur for systems such
as spin-polarized hydrogen, positronium gas, and excitons in a semiconductor.
\end{abstract}
\pacs{PACS numbers: 05.30.Fk, 05.70.Fh, 52.25.Kn, 71.25.+z}

The system consisting of equal numbers of positively and negatively charged
spin-1/2 particles provides a surprisingly rich subject for the study of phase
transitions.  At high density, the system is the well-known neutral Fermi
plasma.
As density decreases, the system becomes unstable to the formation of bound
complexes which behave as bosons. Under certain conditions at low temperature,
these quasi-bosons can form a Bose condensate.  Presently, experimental
searches are under way or proposed for observation of Bose condensation in
several systems of this type, namely excitons in semiconductors
\cite{wolfe-snoke}, hydrogen \cite{greytak,silvera}, and
positronium \cite{platzmann}. Depending on the mass ratio and spin
degeneracies of the positive and negative particles, other phases such as solid
and Fermi liquid can arise.

The phase transition from conducting plasma to insulating bound-pair gas, known
as the Mott transition, has been discussed in the literature for well over
three decades. The general condition for this transition is that the screening
length in the plasma be comparable to or less than the intrinsic
bound-state Bohr radius. A point that has been less well appreciated is the
possibility for {\em bistability and hysteresis} in this transition. The basic
reason for this hysteresis is that neutral bound states will not contribute to
long-range screening. Therefore, if the gas is initially in the insulating
phase, a transition to conducting plasma will occur only when the number of
free
charged particles {\em created by thermal dissociation of bound pairs} becomes
high enough that the screening due to these particles causes bound states to
become unstable. In general, the density of this transition, called the
``ionization catastrophe,''
\cite{rice} is not the same as the Mott transition in the reverse direction
that
occurs when the system is initially in the plasma state.

In this paper, we present a simple model that illustrates the hysteresis in
this
kind of system. Much of the discussion is in terms of equal-mass positive and
negative particles (excitons or positronium) but the discussion can be
generalized to apply to the large-mass-ratio (hydrogen atom) case as well.
For the sake of discussion, the bound states will be called excitons.
In all of the following we assume nondegenerate (Maxwell-Boltzmann) statistics
for both bound states and free particles. This will limit the regions of
phase space where the model is valid (i.e. it will not be valid near the
Bose-Einstein phase transition or for the degenerate Fermi gas), but
the primary effect of hysteresis does not depend on the statistics. We also
ignore van der Waals interactions between bound states and the formation of
bound-state complexes.

{\bf Rate equations}. Instead of the usual partition function $Z =
e^{\Delta/k_BT}$, the ``Planck-Larkin'' partition function
\cite{ebeling,ebeling2},
$Z = e^{\Delta/k_BT} -(1+\Delta/k_BT)$, is used, which takes into account the
fact that states with $\Delta \ll k_BT$ are quasi-free. This first-order
correction to the Boltzmannian partition function was initially calculated in
order to keep the density of states of atoms near the $n \rightarrow \infty$
continuum from an unphysical infinity.  In the context of the neutral plasma
system studied here, it enforces the physical result that when the binding
energy is zero, the population in bound states must become identically zero.
Ebeling et al.\cite{ebeling} have used first-order perturbation theory to show
that this partition function has sound physical basis.

For species conversion via three-body collisions, the rate of change of
the bound state (exciton) population $n_{ex}$ is then given as follows:
\begin{equation}
\frac{\partial n_{ex}}{\partial t} = A n_e \left( n_e^2 - \frac{n_{ex}n_Q}
{e^{\Delta/k_BT} - (1  + \Delta/k_BT)}  \frac{g_eg_h}{g_{ex}}
\left(\frac{\mu}{m}\right)^{3/2} \right)\label{rate}
\end{equation}
where $n_{e}$ is the free electron density, with $n_{ex} + n_e = n$, the
total pair density;  $g_e$, $g_h$ and $g_{ex}$ are the electron, ion (hole) and
exciton spin degeneracy, respectively, and $m$ and $\mu$ are the total and
reduced mass, respectively.
The ``quantum density of states'' is given by
\begin{equation}
n_Q = \left(\frac{mk_BT}{2\pi\hbar^2}\right)^{3/2} \equiv 1/\lambda_D^3
\end{equation}
where $\lambda_D$ is the deBroglie wavelength. Setting $\partial
n_{ex}/\partial t = 0 $ in (\ref{rate}) with $\Delta \gg k_BT$ gives the
standard Saha equation \cite{ebeling} $ (n_e^2/n_Q) = n_{ex}e^{-\Delta/k_BT}
(g_eg_h/g_{ex})(\mu/m)^{3/2}$. The constant $A$, which depends on the cross
section and may be temperature dependent, determines the absolute rate of
conversion and does not enter into steady-state calculations.

The interesting dynamics in this system come from the dependence of $\Delta$ on
$n_e$.  The binding energy (Rydberg) of a single  bound state (exciton) is
given
by $\Delta_0 = \hbar^2/2a^2\mu$, with the Bohr radius $a =
\hbar^2\epsilon/e^2\mu$. (The dielectric constant $\epsilon$ is included in
the case of a polarizable medium, e.g. for excitons in a semiconductor.)  When
screening is present, so that the particles do not interact with pure Coulomb
attraction but with a Yukawa-type interaction, the binding energy decreases
approximately according to
\cite{smith}
\begin{equation}
\Delta(n_e,T) = \left\{ \begin{array}{ll}
\Delta_0 \left( 1 - \frac{2}{1+(qa)^{-1}}\right) & qa<1\\
0 & qa \ge 1. \end{array} \right.
\end{equation}
For $1/q \sim a$,
variational calculations give slight (20\%) corrections to the above
formula \cite{gomes} which are not of interest here.
The screening constant is given by the Debye-Huckel formula,\cite{mahan}
\begin{equation}
q^2 = \frac{4\pi e^2n_e}{\epsilon k_BT}. \label{debye}
\end{equation}
It is important to recognize that only free charged particles contribute to
screening-- the bound states are neutral.

{\bf Numerical results.}
It is convenient to set $n_e=ny^2$ and use $\dot{n}_{ex}=-\dot{n}_{e}$ to
write the dynamics in dimensionless form
\begin{equation}
\frac{dy}{d\tau}=\frac{y}{f(y)}\left[\alpha(1-y^2)-y^4f(y)\right]
\end{equation}
where $\tau=An^2t/2$ is a scaled time, and $f(y)=\exp(F(y))-(1+F(y))$ is the
partition function in terms of
\begin{equation}
F(y)=\mbox{\rm max}\left[0,\frac{(g_eg_h/g)^2}{\pi\beta^2}
\left(\frac{\sqrt{\alpha/\beta}-y}{\sqrt{\alpha/\beta}+y}\right)\right].
\end{equation}
The parameters $\alpha$ and $\beta$ determine the temperature and density
\begin{eqnarray}
\left(\frac{a}{\lambda_D}\right)^2&=&
\frac{m}{\mu}\left(\frac{g}{2g_eg_h}\right)^2\beta^2\\
na^3&=&\left(\frac{g}{g_eg_h}\right)^2\left(\frac{\beta^3}{8\alpha}\right).
\end{eqnarray}

There is an equilibrium at $y=0$ which is unstable due to thermal
ionization of excitons. In addition, there is always at least one stable
equilibrium described by
\begin{equation}
0=\left[\alpha(1-y^2)-y^4f(y)\right]. \label{equil}
\end{equation}
For appropriate choices of $n$ and $T$ there can be three solutions; two
stable and one unstable.  The locus of points $(n,T)$ defining the region of
bistability has been determined numerically. When there is a fully ionized
plasma state at $y=1$ then $F(1)=0$, and the model fails to quantitatively
describe the approach to $y=1$ due to the divergence of $f(y)^{-1}$ as
$F\rightarrow0$.

Figure 1 shows numerical results when $g_e = g_h = 2$ and $g_{ex} = 1$.  The
unitless pair density $na^3$ and the unitless temperature
$(\mu/m)(a^2/\lambda_D^2)$ are control parameters; only the fraction of free
carriers $y^2 = n_e/n$ can vary.  The binding energy $\Delta(n_e)$ is
calculated
using a self-consistent value of
$n_e$. The solid line marks the boundary between the region with {\em one}
steady-state solution and the region with {\em three} steady-state solutions,
one of which is unstable.

Figure 2 illustrates the behavior of the system in these regions, by plotting
the ionized fraction as a function of total pair density at fixed temperature.
For temperature well above the three-solution region (Fig. 2(a)), the system
remains mostly ionized at all densities. At lower temperatures
(Figs. 2(b) and (c)), a sharp transition appears between a pure plasma and
exciton-rich phase. Finally, when temperature is lowered further, a classic
hysteresis curve occurs (Fig. 2(d)). Two different solutions for the ionized
fraction are stable against fluctuations.

{\bf Regions of phase space.} By simple arguments, we can see how hysteresis
occurs in this system. Figure 3 shows the relevant crossovers for the case
$m=4\mu$ and $g_e = g_h = 2, g_{ex} = 1$ of Figs. 1 and 2.  If we assume
that the
system is in the plasma state, then we expect a Mott transition
when $1/a = q$ where $q$ is given by the Debye formula (\ref{debye}) with
$n_e = n$. In terms of the unitless parameters
$na^3$ (unitless pair density) and $a^2/\lambda_D^2$ (unitless temperature),
this becomes
\begin{equation}
na^3 = \frac{1}{2} \left(\frac{\mu}{m}\right) a^2/\lambda_D^2, \label{Mott}
\end{equation}
shown as the heavy solid line in Figure 3.  On the other hand, if we assume the
system is in the insulating gas (excitonic) phase, we would expect a transition
when $1/a = q$ where
$q$ is given by (\ref{debye}), but instead of setting $n_e$ equal to $n$,
we would use the equilibrium value from (\ref{rate}), assuming $\Delta =
\Delta_0$ and $n_e \ll n$. This yields
\begin{equation}
na^3 = \frac{1}{4}
\left(\frac{a^2}{\lambda_D^2}\right)^{1/2}\left(\frac{g}{g_eg_h}\right)
\left(\frac{\mu}{m}\right)^{1/2}
e^{(\lambda_D^2/a^2)(m/\mu)/4\pi},\label{catastrophe}
\end{equation}
shown as the light solid line in Figure 3.  Since at very low temperatures the
number of thermally ionized particles becomes extremely small, the ``ionization
catastrophe'' density becomes exponentially large at low temperature. For
comparison, the conditions for a transition to a Fermi degenerate gas ($na^3
\sim 1$) and for a transition to a weakly-interacting Bose condensate $(na^3 =
2.612(a^2/\lambda_D^2)^{3/2})$ are shown as the dashed and dotted lines in
Figure 1, respectively.  Since the model presented here involves only
Maxwellian
statistics, it is only valid for densities well below both of these curves. The
presence of the plasma state will prevent Bose condensation, however, so the
proximity of the Mott transition to the Bose condensation boundary in certain
regions of phase space merits attention in all composite-boson condensation
experiments.

At low temperature and low density, the system is an insulating
excitonic (bound-state) gas, assuming repulsive interactions between excitons.
As density is raised, in the low-temperature limit the system remains an
insulating gas until an ionization catastrophe occurs. As seen in Fig. 3, this
can occur at substantially higher density than the Mott transition given by Eq.
(\ref{Mott}). If the system starts in the plasma state in this temperature
regime, however, it will remain a plasma indefinitely unless the density falls
below the Mott density given by (\ref{Mott}).

These two curves essentially form the boundaries of the region of hysteresis
shown in Fig. 1. (The curve in Fig. 3 given by (\ref{catastrophe}) lies above
the phase boundary for an ionization catastrophe found numerically in
Fig. 1 because it does not use a self-consistent value of the binding energy
$\Delta(n_e)$.) At temperatures above the intersection of the solid curves,
however, the system undergoes a single phase transition to a plasma state,
since
an ionization catastrophe occurs at lower density than the Mott transition
given
by (\ref{Mott}). At very high temperature, the ``ionization catastrophe'' is
not a clearly defined transition, since even at low density a substantial
fraction of the system is ionized.

{\bf Application to experiment.} Preliminary results from experiments with
excitons in the semiconductor Cu$_2$O seem consistent with this picture. As
reported in Ref. \cite{shields}, excitons with binding energy $\Delta=150$ meV
exist in this material and have been observed at room temperature. In those
experiments, the value of $na^3$ was roughly 0.003 (pair density $\sim 10^{19}$
cm$^{-3}$, excitonic Bohr radius $7 \AA$), $a^2/\lambda_D^2$ was approximately
0.04 (T = 300K, mass $m = 3m_0$), and $\mu/m \simeq 4$ (roughly equal electron
and hole mass.) This puts the system right in the region of the crossover seen
in Figures 1 and 3. In the experiments, excitons appeared only when
created resonantly by laser excitation. When free carriers were created by
laser
excitation with photon energy well above the band gap, no exciton luminescence
ever appeared. In other words, two different stable states occurred at roughly
the same temperature and pair density.

Clearly, there is room for more experiments and theory on this interesting
system. Since the model presented here assumes spatial homogeneity, pattern
formation and related nonlinear dynamical effects do not occur. The possibility
exists that when the effects of spatial variations are taken into account,
regions of coexistence and very interesting chaotic behavior may be seen in the
exciton/plasma phase transitions.

These results also relate to experiments on phase transitions of
atoms with attractive interactions, e.g. the search for a metallic
phase of hydrogen at high pressure \cite{hanfland}. Because of attractive
interactions, the lower, insulating state may not be a {\em gas}, as assumed
here, but instead a solid or liquid state. In many cases, however, screening is
assumed to play an important role in a transition to a conducting phase. The
results here indicate that the densities necessary for a {\em spontaneous}
transition to such a phase may be much higher than expected from equilibrium
calculations, since a metastable solid or liquid state comprised of bound pairs
will have poor screening even at high density.

\begin{figure}

FIG. 1.  The boundary between the regions of phase space with
one steady-state solution and three steady-state solutions of Eq. (\ref{equil})
in the text, for the case $g_e = g_h = 2$, $g=1$.

\end{figure}

\begin{figure}

FIG. 2. Steady-state solutions of $n_e/n$ from Eq. (\ref{equil}), as a function
of $na^3$, for four temperatures.

\end{figure}

\begin{figure}

FIG. 3. The regions of phase space indicated by the simple equations
(\ref{Mott}) and (\ref{catastrophe}) in the text. The model breaks down above
the critical density for Bose condensation (dotted lines, for two different
values of $m/\mu$ corresponding to excitons and hydrogen) and in the Fermi
degenerate region ($na^3 \sim 1$, indicated by the dashed line.)
\end{figure}

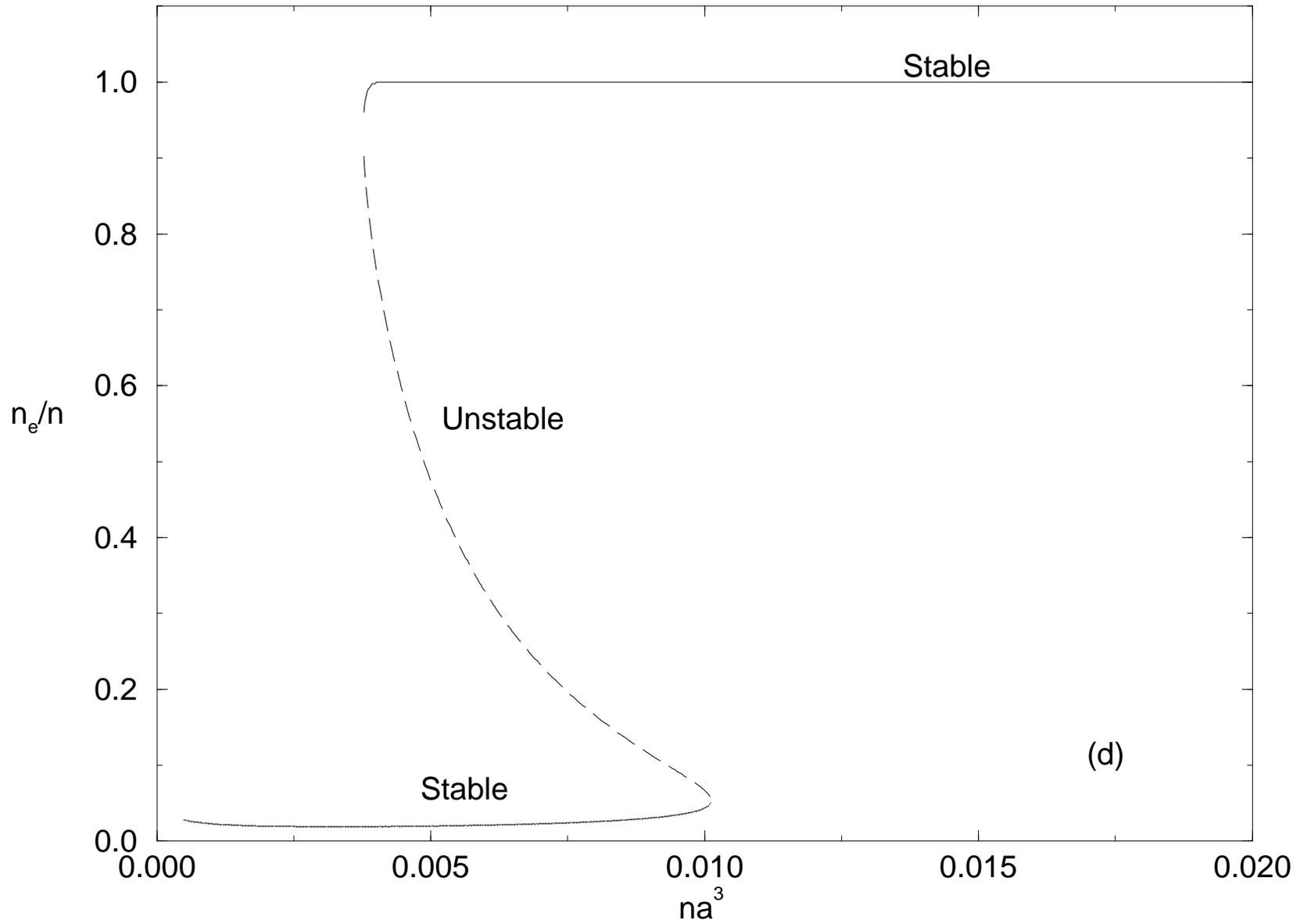
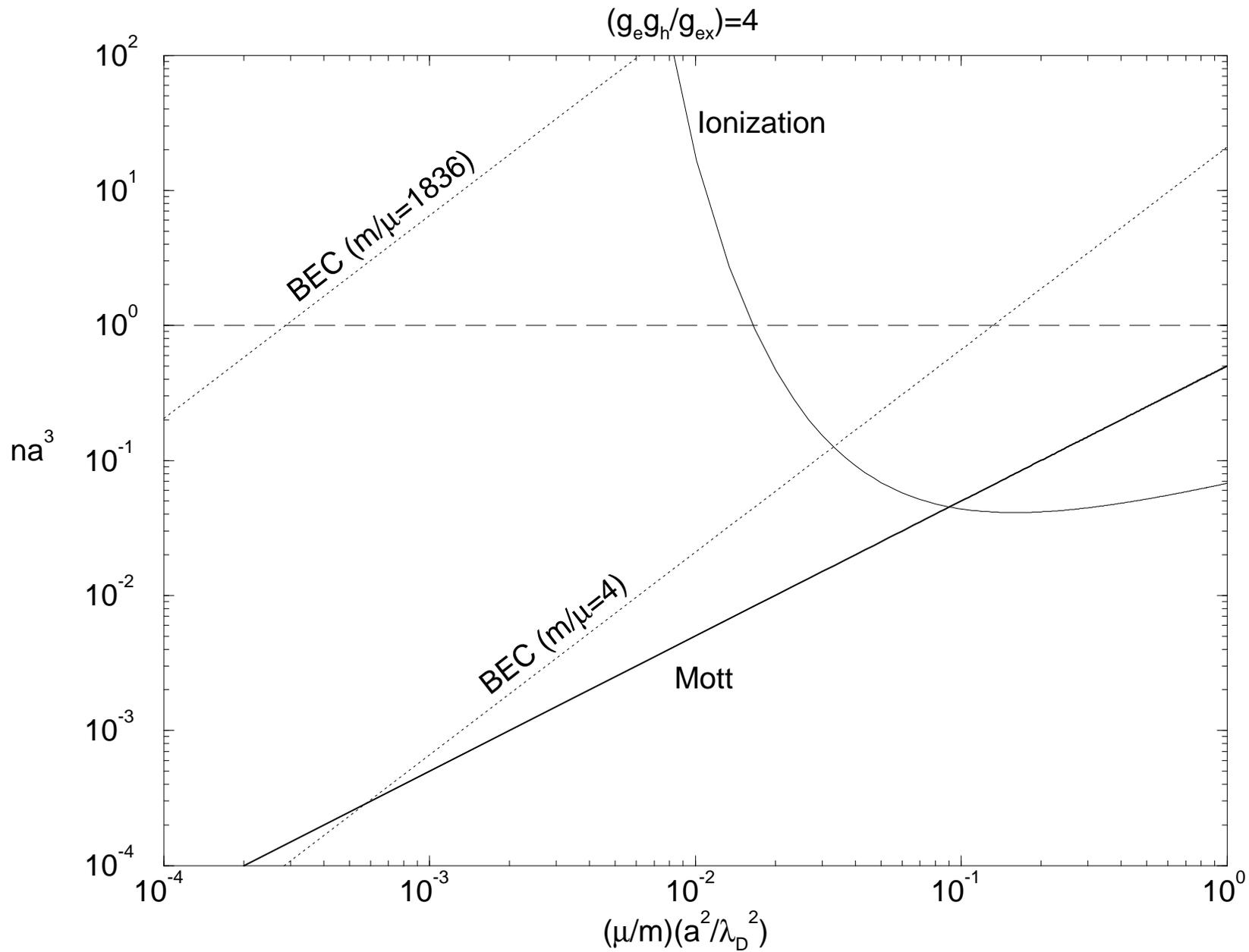
\end{document}